\newif\ifcqg\cqgfalse
\ifcqg
\documentclass[]{iopart}
\usepackage{iopams}
\usepackage{setstack}
\eqnobysec
\def\flamoi{\fl}

\else
\documentclass[twocolumn,floatfix,preprintnumbers,aps,prl,showpacs,footinbib,amsmath,amssymb,amsfonts]{revtex4}
\usepackage{bm}

\def\flamoi{}

\fi

\usepackage[dvips,usenames]{color}
\usepackage{dsfont}
\usepackage{hyperref}

\newcommand{\be}{\begin{equation}}
\newcommand{\ee}{\end{equation}}
\newcommand{\bea}{\begin{eqnarray}}
\newcommand{\eea}{\end{eqnarray}}
\newcommand{\gr}[1]{\mathbf{#1}}

\newcommand{\ii}{\mathrm{i}}

\newcommand{\frd}[2]{\frac{\dd #1}{\dd #2}}

\newcommand{\eb}{\bar{e}}

\newcommand{\uline}[1]{\underline{#1}}
\def\st{\sigma_{\mathrm T}}
\def\me{m_{\mathrm e}}

\def\dd{{\rm d}}
\def\HH{\mathcal{H}}

\definecolor{Myblue}{rgb}{0,0,1}
\definecolor{Myred}{rgb}{1,0,0}
\def\zB{{\color{Myred}O}}
\def\iB{{\color{Myred}I}}
\def\jB{{\color{Myred}J}}
\def\kB{{\color{Myred}K}}

\def\aB{{\color{Myred}A}}
\def\bB{{\color{Myred}B}}
\def\cB{{\color{Myred}C}}

\def\zT{{{\color{Myblue}o}}}
\def\iT{{{\color{Myblue}i}}}
\def\jT{{{\color{Myblue}j}}}
\def\kT{{{\color{Myblue}k}}}
\def\lT{{{\color{Myblue}l}}}
\def\mT{{{\color{Myblue}m}}}
\def\pT{{{\color{Myblue}p}}}
\def\qT{{{\color{Myblue}q}}}
\def\aT{{{\color{Myblue}a}}}
\def\bT{{{\color{Myblue}b}}}
\def\cT{{{\color{Myblue}c}}}
\def\dT{{{\color{Myblue}d}}}

\def\hT{{{\color{Myblue}h}}}
\def\zTt{{\color{Myblue}{\tilde o}}}
\def\iTt{{\color{Myblue}{\tilde \imath}}}

\def\aTt{{\color{Myblue}{\tilde a}}}

\def\ir{{\mathrm r}}
\def\ie{{\mathrm e}}

\def\hashamoi{\#}

\begin{document}
\title[The radiative transfer for polarized radiation at second order in
cosmological perturbations]{The radiative transfer for polarized radiation at second order in
cosmological perturbations}
\author{Cyril Pitrou}
\ifcqg
\ead{cyrilp@astro.uio.no}
\address{Institute of Theoretical Astrophysics,
         University of Oslo, 
         P.O. Box 1029 Blindern, 0315 Oslo, Norway.}
\else
\email{cyrilp@astro.uio.no}
\affiliation{Institute of Theoretical Astrophysics,
         University of Oslo, 
         P.O. Box 1029 Blindern, 0315 Oslo, Norway.}
\fi


\date{\today}

\begin{abstract}
This article investigates the full Boltzmann equation up to
second order in the cosmological perturbations. Describing the distribution of
polarized radiation by using a tensor valued distribution function, the second order Boltzmann equation, including
polarization, is derived without relying on the Stokes parameters. 
\end{abstract}
\ifcqg
\else
\maketitle
\fi
\subsection{Introduction}

The properties of the cosmic microwave background (CMB) temperature
fluctuations depend on both the initial conditions set at the end
of the primordial inflationary era, and on their post-inflationary evolution. We describe radiation using the
kinetic theory, in which its properties are encoded in a distribution
function. The theory of cosmological perturbations around a maximally symmetric space-time enables to solve order by order the
evolution equation for the CMB anisotropies. The linear perturbations fail to capture the
intrinsic non-linear features of General Relativity which can enter both the
initial conditions and the evolution, and we thus need at least the second
order if we want to estimate the bispectrum in the CMB and better reveal the physics of the primordial universe. Indeed, the
bispectrum can only be generated either from non-Gaussian initial conditions
set by inflation or by secondary effects during the subsequent evolution, and
though standard scalar field slow-roll inflation predicts negligible amounts
of non-Gaussianity~\cite{Maldacena2003}, the latest observations of the CMB~\cite{Komatsu2008,Smith2009}
indicate that there might be a non-vanishing bispectrum. We thus need
to extend the program followed at first order in perturbations up to second order, that is to build a full set of second order gauge invariant perturbation variables and derive the perturbed
Boltzmann and Einstein equations which determine their dynamics. In the fluid limit, the gauge issue was studied in Ref.~\cite{Bruni1997} and gauge invariant variables
were built in Refs.~\cite{Malik2004,Nakamura2007} up to second order in perturbations. This
fluid approximation has already been used to understand the general form of the
bispectrum on small scales generated by evolutionary effects in Ref.~\cite{PUB2008}. As for the
kinetic theory, the gauge issue was studied in our previous
paper~\cite{Pitrou2007}, and the evolution equations through free-streaming
and Compton collision with free electrons were derived up to second order
in Refs.~\cite{Dodelson1993,Hu1994,Bartolo2006,Maartens1999}, but it is so far restricted to
unpolarized radiation, which is inconsistent since Compton scattering does generate polarization. This letter summarizes the full second order
derivation of the radiation transfer, including polarization, detailed in Ref.~\cite{Pitrou2008}.

\subsection{Kinetic theory with polarization}

{\it Describing polarized radiation} The momentum of photons is usually decomposed on an orthonormal basis, that is using a tetrad field
defined by $\gr{e}_{\aT}.\gr{e}_{\bT}\equiv e_{\aT}^{\,\,\mu} e_{\bT}^{\,\,\nu} g_{\mu
  \nu}=\eta_{\aT \bT}$, $\gr{e}^{\aT}.\gr{e}^{\bT}\equiv e^{\aT}_{\,\,\mu} e^{\bT}_{\,\,\nu}
g^{\mu \nu}=\eta^{\aT \bT}$.
We use Greek indices ($\mu,\nu,\rho\dots$) for abstract indices and the
beginning of the Latin alphabet ($\aT,\bT,\cT\dots=0\dots3$) for tetrad labels. We also use $\iT,\jT,\kT\dots=1\dots3$ for the spatial type tetrads and we reserve the label $\zT$ for
the timelike vector. As for the labels associated with the coordinates, we use $\aB,\bB,\cB,\dots=0\dots3$, and $\iB,\jB,\kB,\dots=1\dots3$, with
the time component labeled by $\zB$.\\
A momentum can then be written $\gr{p}=p^\aT \gr{e}_\aT=p^\zT \gr{e}_\zT
+p^\iT \gr{e}_\iT$, and decomposed between the energy $p^\zT$ and the
spacelike direction unit vector $\gr{n}$ according to $p^\mu=p^\zT (e_{\zT}^{\,\mu} +n^\mu)$ with $n_\mu e_{\zT}^{\,\mu}=0$.
This decomposition can be used to define the screen projector
\be\label{DefScreen}
{S}_{\mu\nu}(\gr{p})=g_{\mu \nu}+e^\zT_{\,\mu} e^\zT_{\,\nu}-n_\mu n_\nu\,, 
\ee
which projects on the space both orthogonal to $\gr{e}_\zT$ and to the direction $\gr{n}$. The radiation is represented by a Hermitian tensor valued distribution function
also called polarization tensor~\cite{Challinor2000b,Tsagas2007} satisfying $p^\mu F_{\mu\nu}(p^\aT)=0$.
The screen-projected distribution function, $f_{\mu\nu}(p^\aT)=S_\mu^\rho S_\nu^\sigma F_{\rho \sigma}(p^\aT)
$, has four degrees of freedom which can be split according to
\be\label{Decf}
 f_{\mu \nu} \equiv \frac{1}{2} IS_{\mu  \nu}+P_{\mu \nu}
+\frac{\ii}{2} V \epsilon_{\mu \nu \sigma}
n^\sigma,
\ee
where $\epsilon_{\mu \nu \sigma}\equiv e_\zT^\rho\epsilon_{\rho \mu \nu
  \sigma}$, with $\epsilon_{\rho  \mu \nu \sigma}$ the space-time fully
antisymmetric tensor, and where we have omitted for simplicity of notation the dependence in
$p^\aT$. $P_{\mu \nu}$, which encodes the degree of linear polarization, is real, symmetric and trace free, as well as orthogonal to
$\gr{e}_\zT$ and $\gr{n}$. It has two degrees of freedom, usually described by the Stokes functions $Q$ and $U$. $I$ and $V$ are respectively the intensity (or distribution function) for both polarizations and the circular polarization. The brightness is then defined as
\be\label{brightness}
{\cal I}(n^\iT)\equiv \frac{4 \pi}{(2 \pi)^3} \int
I(p^\zT,n^\iT)(p^\zT)^3 \dd p^\zT\,,
\ee
and we can define similarly ${\cal V}$ and ${\cal
  P}_{\mu \nu}$.\\
The remaining dependence in $\gr{n}$, can be further expanded in multipoles using
projected symmetric trace-free (PSTF) tensors, where projected means that they
are orthogonal to $\gr{e}_\zT$. For instance, using the notation $n^{\uline{\iT_\ell}}\equiv n^{\iT_1}\dots
n^{\iT_\ell}$ and $n^\iT\equiv n^\mu e^{\iT}_\mu$, ${\cal I}$ can be expanded in
\be\label{Def_multipoleS}
{\cal I}(n^\iT)=\sum_{\ell=0}^\infty {\cal I}_{\uline{\iT_\ell}}n^{\uline{\iT_\ell}}\,,
\ee
where the ${\cal I}_{\uline{\iT_\ell}}\equiv {\cal I}_{\iT_1\dots\iT_\ell}$
are PSTF, and where conventionally, the lowest multipole, i.e. corresponding to $\ell=0$, is noted ${\cal I}_\emptyset$. A similar expansion can
be performed on ${\cal V}$. As for ${\cal P}_{\aT \bT}\equiv {\cal P}_{\mu
  \nu} e_\aT^{\,\mu}e_\bT^{\,\nu}$, its non vanishing components can be expanded in electric and magnetic type multipoles according to~\cite{Challinor2000b,Tsagas2007}
\be\label{Def_multipoleP}
\flamoi {\cal P}_{\iT \jT}(n^\aT)=\sum_{\ell=2}^\infty \left[{\cal E}_{\iT \jT
  \uline{\kT_{\ell-2}}}n^{ \uline{\kT_{\ell-2}}} \,- n_{\mT}\epsilon^{\mT
  \lT}_{\,\,\,\,\,(\iT}{\cal B}_{\jT)\lT \uline{\kT_{\ell-2}}}n^{\uline{\kT_{\ell-2}}}\right]^{\mathrm{TT}}
\ee
where the notation $\mathrm{TT}$ denotes the transverse (to $\gr{n}$)
symmetric trace-free part. We have used $(..)$ for the symmetrization of indices, and we
will also use $[..]$ for the antisymmetrization.\\

{\it Transformation rules under a change of frame} This description depends on the tetrad chosen to define the polarization
tensor, the energy and the momentum direction. Under a local boost
parameterized by $v^\iT \equiv  v^\mu e^{\iT}_\mu$, the tetrad transforms to $\tilde e_\aT = e_\bT \Lambda^\bT_{\,\,\aT}$,
where the components of the transformation are given by
$\Lambda^\zT_{\,\zT}=\gamma\equiv(1-\gr{v}.\gr{v})^{-\frac12}$, $\Lambda^\zT_{\,\iT}=-\gamma v_\iT$ and
$\Lambda^\iT_{\,\jT}=\delta^\iT_\jT+[\gamma^2/(\gamma+1)]v^\iT
v_\jT$. Thus, the magnitude and the direction unit vector of the photon momentum transform to
\bea\label{Eq_Tmomentum1}
p^{\zTt}&\equiv& \gr{p}.\tilde{\gr{e}}^{\zT}=\gamma p^\zT\left(1-\gr{n}.\gr{v} \right)\,,\\*
\tilde n^{\iTt}&\equiv&{\tilde{\gr{n}}}. {\tilde{\gr{e}}}^\iT=\frac{1}{\gamma(1-\gr{n}.\gr{v})}\left[n^\iT+\frac{\gamma^2}{(1+\gamma)}\gr{n}.\gr{v} \,v^\iT-\gamma v^\iT \right].\nonumber
\eea
This implies that
\bea\label{TruleforS}
\tilde S_{\mu \nu}=\tilde S_\mu^{\,\,\rho} \tilde S_\nu^{\,\,\sigma} S_{\rho
  \sigma}\,,\qquad \tilde n^\mu \tilde \epsilon_{\mu \rho \sigma}=n^\mu
\epsilon_{\mu \alpha \beta} \tilde S^\alpha_{\,\,\rho}\tilde S^\beta_{\,\,\sigma}\,.
\eea  
Since the screen-projected polarization tensor transforms to
\be\label{Eq_Tpropertydebase}
\tilde f_{\mu \nu}(p^{\zTt},\tilde n^\aTt)=\tilde S_\mu^{\,\,\alpha} \tilde S_\nu^{\,\,\beta}
f_{\alpha \beta}(p^{\zT},n^\aT)\,,
\ee
then it can be checked from the decomposition~(\ref{Decf}) and the transformation
rules~(\ref{TruleforS}) that $I$ and $V$ transform as scalars, that is $\tilde
I(p^{\zTt},\tilde n^\aTt)=I(p^{\zT},n^\aT)$, $\tilde V(p^{\zTt},\tilde
n^\aTt)=V(p^{\zT},n^\aT)$, and that $P_{\mu\nu}$ transforms as $f_{\mu\nu}$. The differential solid angle of the momentum direction transforms according to $\dd \tilde \Omega =\left[{\gamma(1-\gr{v}.\gr{n})}\right]^{-2}\dd \Omega$,
and thus the transformation rules of the energy integrated multipoles of the
brightness can be deduced to be given by
\be\label{Eqinttochangeframe}
\flamoi\tilde{\cal I}_{\uline{\tilde{\iT}_\ell}}=\frac{1}{\Delta_\ell}\int \dd
\Omega \left[\gamma(1-\gr{v}.\gr{n})\right]^{2}\sum_{\ell'=0}^\infty {\cal I}_{\uline{\jT_{\ell'}}}n^{\uline{\jT_{\ell'}}} \tilde n_{\langle \uline{\tilde{\iT}_\ell}\rangle},
\ee
where $\Delta_\ell \equiv (4 \pi \ell!)/[(2\ell+1)!!]$ and
$\langle\dots\rangle$ is the notation for the symmetric trace-free part. The transformation rules of the electric and magnetic multipoles are detailed in Refs.~\cite{Pitrou2008,Tsagas2007}.\\

{\it The Boltzmann equation} The evolution of the polarization tensor is dictated by the Boltzmann equation
\be\label{EqBoltzmann}
L[f_{\aT \bT}]\equiv S_\aT^{\,\cT}
S_\bT^{\,\dT}\left[p^\hT \nabla_\hT f_{\cT
    \dT}+ \frac{\partial f_{\cT
      \dT}}{\partial p^\hT}\frac{\dd p^\hT}{\dd s}
\right]=C_{\aT \bT}\,,
\ee
where $L[]$ is the Liouville operator and $C_{\aT \bT}$ the collision tensor. It can be shown that the Liouville operator preserves the decomposition
of $f_{\mu \nu}$ in an antisymmetric part ($V$), a trace ($I$) and a symmetric
traceless part ($P_{\mu \nu}$), that is
\be\label{Liouvillesepcomponents}
\flamoi L[f_{\aT \bT}]=\frac{1}{2}L[I]S_{\aT \bT}+L[P_{\aT
  \bT}]+\frac{\ii}{2} L[V]n^\cT \epsilon_{\cT \aT \bT}\,.
\ee 

\subsection{Cosmological perturbations}

{\it Perturbation of the metric} We assume that, at lowest order, the universe is well described by a
Friedmann-Lema\^{\i}tre space-time (FL) with Euclidian spatial sections. The most
general form of the metric for an almost FL universe is
\be\label{metric}
\flamoi \dd s^2 = a(\eta)^2 \left[-(1 + 2\Phi )\dd\eta^2 + 2
 \omega_{\iB} \dd x^{\iB}\dd\eta +  h_{\iB\jB}\dd x^{\iB}\dd x^{\jB}\right],\nonumber
\ee
where $\eta$ is the conformal time for which the corresponding index is $\zB$,
and $a(\eta)$ is the scale factor. We define the Hubble parameter by
$\HH\equiv a'/a$ where a prime denotes a derivative w.r.t. $\eta$. We work in
the Poisson gauge though, as already mentioned, a complete treatment of perturbations can be performed by defining gauge-invariant variables~\cite{Nakamura2007,Pitrou2007}, and we perform a scalar-vector-tensor decomposition in $\omega_{\iB}=\partial_{\iB} B + B_{\iB}$, $h_{\iB \jB}=(1-2 \Psi)\delta_{\iB\jB} +2 H_{\iB\jB}$, where $B_\iB$ and $H_{\iB \jB}$ are transverse ($\partial^\iB B_\iB=\partial^\iB H_{\iB \jB}=0$), and $H_{\iB \jB}$
is traceless ($H^{\iB}_{\,\,\iB}=0$). Each of these variables can be split in first and second-order
parts as $  W=W^{(1)}+\frac{1}{2}W^{(2)} $, and if a quantity has also a background
value it is noted $\bar W$. First-order variables are solutions of first-order
equations, whereas second-order equations will involve purely second-order terms, e.g.
$W^{(2)}$ and terms quadratic in the first-order variables, e.g.
$[W^{(1)}]^2$ for which we will omit the order superscript. We neglect the first
order vector modes ($B_\iB^{(1)}=0$) which decay and the first order tensor
modes ($H_{\iB\jB}^{(1)}=0$) which are expected to be very small
in standard models of inflation.\\

{\it Perturbations of tetrads and Ricci rotation coefficients} The perturbed tetrad can be expressed in function of the background tetrad in
the generic form by $\gr{e}^{(n)}_\aT =R^{(n)\bT}_{\aT} \gr{\eb}_\bT$,
$\gr{e}^{\bT(n)} = \gr{\eb}^\aT S^{(n)\bT}_{\aT}$. The symmetric part
$R^{(n)}_{(\aT\bT)}$ is constrained by the normalization conditions of the
tetrad and is thus expressed in function of the metric perturbations. We are
free to choose the antisymmetric part $R_{[\aT \bT]}$, since it corresponds to the Lorentz transformation freedom (boost and rotation). We require $\gr{e}^\zT \sim \gr{\dd\eta} $~\cite{Durrer1994}, which is equivalent to choose $R^{(n)}_{\iT \zT}= S^{(n)}_{\iT \zT}=0$ for any $n$. We also fix the rotation by requiring
$R^{(n)}_{[\iT \jT]}=S^{(n)}_{[\iT \jT]}=0$. This procedure means that our tetrad is adapted to observers whose velocity is always orthogonal to
constant time hypersurfaces. We then choose a background tetrad adapted to our
coordinates, i.e. $\bar{e}_{\bT}^{\,\cB}=\delta_{\bT}^{\cB}/a$  and
$\bar{e}^{\bT}_{\,\cB}=a\delta^{\bT}_{\cB}$, which finally enables us to
express the perturbations of the tetrad in function of the metric
perturbations. The perturbations of the Ricci rotation coefficients $\omega_{\aT \bT \cT} \equiv \eta_{\bT \dT
}e^\dT_{\,\,\nu}e_\aT^{\,\,\mu}\nabla_{\mu}e_\cT^{\,\,\nu}$, needed to express
the covariant derivative in the tetrad basis, follows directly from the
perturbation of the tetrad, and the components used in this paper are
$\omega^{(1)}_{\zT\iT\jT}=0$, and 
\bea\label{Pertomega}
\flamoi \omega^{(1)}_{\jT\iT\kT} &=& -\frac{2}{a}\partial_{[\kB} \Psi^{(1)}  \delta_{\iB]\jB}\\
\flamoi \omega^{(2)}_{\zT\zT\iT}&=&\frac{1}{a}\left[-\partial_\iB
  \Phi^{(2)}+2(2\Phi-\Psi)\partial_\iB\Phi\right],\nonumber\\
\flamoi \omega^{(2)}_{\iT\zT\jT}&=&\frac{1}{a}\left[\partial_{(\iB}B_{\jB)}^{(2)}- H^{(2)'}_{\iB\jB}+\left(\HH  \Phi^{(2)}+\Psi^{(2)'}\right)\delta_{\iB\jB}\right. \nonumber\\
\flamoi &&\left.\quad -(3\HH \Phi+2\Psi')\Phi\delta_{\iB\jB} +4
  \Psi\Psi'\delta_{\kB(\jB}\delta^{\kB}_{\,\iB)}\right].\nonumber
\eea

\subsection{The perturbed Liouville operator}

In practice, we want to express the Boltzmann equation in function of $\eta$, since we
want to perform an integration on coordinates. We should thus multiply Eq.~(\ref{EqBoltzmann}) by $\dd s/\dd \eta=1/p^\zB$. However, there is no point multiplying with the full expression of $\dd
  s/\dd \eta$ since it would then bring metric perturbations in the collision tensor. Instead, we multiply only by $\left(\dd s/\dd \eta\right)^{(0)} =
\left(1/p^\zB\right)^{(0)}=a/p^\zT$, and we will use the notation $L^\hashamoi[] \equiv L[ ]\left(\dd s/\dd \eta\right)^{(0)}$ and a similar definition for the collision term. Since we are also exclusively interested in
the dynamics of the brightness, we focus our attention on the energy
integrated Liouville operator ${\cal L}^\hashamoi[]$ defined similarly to
Eq.~(\ref{brightness}), and we use a similar definition for ${\cal C}_{\aT
  \bT}^\hashamoi$. The perturbation of the Liouville operator up to second order is of the form
\bea\label{DecL}
\flamoi &&{\cal L}^\hashamoi[\bar{\cal X},{\cal X}^{(1)},{\cal
  X}^{(2)}]=\bar{\cal L}^\hashamoi[\bar{\cal X}]+{\cal
  L}^{\hashamoi(1)}[\bar{\cal X},{\cal X}^{(1)}]\\
&&\qquad+\frac{1}{2}\left\{{\cal L}^{\hashamoi(2)}[\bar{\cal X},{\cal
    X}^{(2)}]+{\cal L}^{\hashamoi(1)(1)}[\bar{\cal X},{\cal X}^{(1)}] \right\},\nonumber
\eea
where ${\cal X}$ stands for either ${\cal I}$, ${\cal V}$ or ${\cal P}_{\aT \bT}$. A similar decomposition
is performed for the collision term.  Since ${\cal L}^{\hashamoi(2)}[]$
contains the terms linear in purely second order variables, it has the same functional
form as ${\cal L}^{\hashamoi(1)}[]$, and consequently we only need the expressions of ${\cal
  L}^{\hashamoi(2)}[]$ and ${\cal L}^{\hashamoi(1)(1)}[]$ to report the
Boltzmann equation up to second order in perturbations.\\
We first need the perturbation of the trajectory at first order which is $\left(\dd x^\iB/\dd \eta\right)^{(1)}=\left(p^\iB/p^\zB\right)^{(1)}=n^\iT
(\Phi+\Psi)$. Then, from the perturbations of the Ricci rotation coefficients~(\ref{Pertomega}), we deduce the perturbations of the geodesic
equation which are given by
\be\left(\frd{p_\aT}{s}\right)^{(n)}+\omega^{(n)}_{\bT\aT\cT}p^\cT p^\bT=0\,.
\ee
This is used to obtain the second order evolution equation for the energy and
the first order evolution equation for the direction (the lensing equation) which read
\bea\label{Eq_evolution_pi0_ordre_2}
\left(\frd{p^\zT}{\eta}\right)^{(2)}&=&p^\zT\left[-n^\iT \partial_\iB \Phi^{(2)}
  +\left(\partial_\iB B_{\jB}^{(2)}-H_{\iB\jB}^{(2)'}\right) n^\iT n^\jT\right.\nonumber\\
&&\left.+\Psi^{(2)'}+2 (\Phi-\Psi)n^\iT \partial_\iB \Phi + 4\Psi \Psi'  \right],\\
\left(\frd{n^\iT}{\eta}\right)^{(1)}&=&- S^{\iT\jT}\partial_\jB \left(\Psi+\Phi \right)\,.
\eea
This is all what is required to obtain from Eq.~(\ref{EqBoltzmann}) the second
order Liouville operator for ${\cal X}={\cal  I},{\cal V},{\cal P}_{\aT\bT}$
\bea\label{Liouville2}
\flamoi &&{\cal L}^{\hashamoi(2)}[\bar {\cal X}, {\cal X}^{(2)}]=\frac{\partial 
  {\cal X}^{(2)}}{\partial \eta}+ n^\jT \partial_\jB  {\cal X}^{(2)} +4\HH
 {\cal X}^{(2)}\\
\flamoi &&-4\delta_{\cal X}^{\cal I}\,\bar {\cal X}\left[- n^\jT \partial_\jB  \Phi^{(2)} +   \Psi'^{(2)}
  +\left(\partial_\iB  B_{\jB}^{(2)}- H_{\iB\jB}^{(2)'}\right) n^\iT
  n^\jT\right]\,,\nonumber
\eea
\bea\label{Liouville11}
\flamoi &&{\cal L}^{\hashamoi(1)(1)}[\bar {\cal X}, {\cal X}^{(1)}]=2 ( \Phi+ \Psi)n^\iT \partial_\iB
{\cal X}^{(1)} \nonumber\\
\flamoi &&+8{\cal X}^{(1)}\left(n^\iT \partial_\iB  \Phi- \Psi' \right) -8\,\delta_{\cal X}^{\cal I}\bar
 {\cal X} \left[( \Phi- \Psi)  n^\iT \partial_\iB  \Phi + 2  \Psi  \Psi' \right]\nonumber\\
\flamoi &&-2S^{\iT\jT}\partial_\jB  (\Psi+\Phi)\frac{\partial  {\cal
    X}^{(1)}}{\partial n^\iT}-2  \Phi {\cal L}^{\hashamoi(1)}[\bar{\cal X}, {\cal X}^{(1)}]\,,
\eea
where it is implied that for ${\cal X}={\cal P}_{\aT\bT}$, according to
Eqs.~(\ref{EqBoltzmann}) and (\ref{Liouvillesepcomponents}), this expression also
needs to be screen projected. The notation $\delta_{\cal X}^{\cal I}$ has also
been introduced in terms involving $\bar{\cal X}$ to remind that, due to the
symmetries of the background space-time, there is no circular or linear polarization at the background level.

\subsection{The collision term}

{\it Electrons without bulk velocity} We consider the case where the free electrons have only a bulk velocity and no thermal dispersion in their
velocity distribution. Since all electrons have the same velocity, we choose to align the first vector of the
tetrad $\gr{e}_\zT$ with this bulk velocity, that is to work in the baryons rest frame. In that case, we can neglect the terms
arising from the recoil of electrons, since they will be of order $T_\ir/\me
\sim 10^{-6}$ around recombination, where $T_\ir$ is the temperature of
radiation, and consequently they do not contribute to the bispectrum. The Thomson approximation is thus
sufficient and, in this case, the collision tensor is known and reads~\cite{Portsmouth2004,Tsagas2007} 
\be
C^\hashamoi_{\iT \jT}(p^\aT)=  \tau' \left[\frac{3}{2}\int \frac{\dd \Omega'}{4 \pi}
  S_{\iT}^{\,\kT} S_{\jT}^{\,\lT} f_{\kT \lT}(p'^\aT) -f_{\iT \jT}(p^\aT)\right]\,,
\ee
where $\tau'=a n_\ie \st$, with $n_\ie$ the electrons number density, and
$\st$ the Thomson cross section. Explicitly, for the energy integrated collision
term we obtain
\bea\label{CollisionThomson}
\flamoi &&\frac{{\cal C}^\hashamoi_{\iT \jT}(n^\qT)}{\tau '}=\left[-{\cal P}_{\iT \jT}(n^\qT)-\frac{1}{10}{\cal I}_{\iT \jT}
  +\frac{3}{5}{\cal E}_{\iT \jT} \right]^{\mathrm{TT}}\nonumber\\
\flamoi &&+\frac{1}{2}S_{\iT
  \jT}\left[- {\cal I}(n^\qT)+{\cal I}_\emptyset+\frac{1}{10}{\cal I}_{\kT\lT}n^\kT
  n^\lT-\frac{3}{5}{\cal E}_{\kT \lT}n^\kT n^\lT\right]\nonumber\\
\flamoi &&+\frac{1}{2}\ii \epsilon_{\iT \jT
  \kT}n^\kT\left[-{\cal V}(n^\qT)+\frac{1}{2}{\cal V}_\lT n^\lT \right]\,.
\eea
We clearly see on this expression that the circular polarization is not
excited and thus remains null if so initially. This explains why we did not bother reporting the corresponding part of the Liouville operator,
and from now on we neglect it. Now, if the electrons have a thermal velocity,
which is in fact the case, the previous expressions has corrections of order
$T_\ie/m_\ie\simeq T_\ir/m_\ie$ where $T_\ie$ is the temperature of electrons, but as for the recoil term, we can discard them as they will not contribute to the bispectrum.\\

{\it Electrons with bulk velocity} If the distribution of electrons has a bulk
velocity relative to the cosmological frame, then we just need to transform the result obtained in the
baryons rest frame to the cosmological frame, using the transformation
rules~(\ref{Eq_Tmomentum1}) (\ref{Eq_Tpropertydebase}) and (\ref{Eqinttochangeframe}) and remembering that the
energy integration is done according to Eq.~(\ref{brightness}). Keeping only the terms which can contribute eventually to the second
order expansion, we obtain 
\bea
\flamoi &&\frac{{\cal C}^\hashamoi_{\iT \jT}(n^\qT)}{\tau'}=\frac{1}{2}S_{\iT \jT}\left[-\frac{1}{2}{\cal I}_\kT n^\kT
\gr{n}.\gr{v}+\frac{1}{2}{\cal I}_{\kT\lT} n^\kT n^\lT \gr{n}.\gr{v}\right.\\
\flamoi &&+\left(
  -{\cal I}(n^\qT)+{\cal I}_\emptyset+\frac{1}{10}{\cal I}_{\kT\lT}n^\kT
  n^\lT\right)\left(1-\gr{n}.\gr{v}\right)-\frac{1}{2}{\cal I}_\kT v^\kT\nonumber\\
\flamoi &&+\frac{1}{5}\left(6{\cal E}_{\kT \lT}-{\cal I}_{\kT\lT}\right) n^\kT v^\lT +4{\cal I}_\emptyset \gr{n}.\gr{v}-{\cal I}_\emptyset\gr{v}.\gr{v}+7{\cal I}_\emptyset(\gr{v}.\gr{n})^2\nonumber\\
\flamoi &&\left.-\frac{3}{5}{\cal E}_{\kT \lT}n^\kT
  n^\lT(1+5\gr{v}.\gr{n})-\frac{6}{5}{\cal B}_{\kT\lT}\epsilon^{\mT\pT\lT}v_{\pT}n_{\mT} n^\kT\right]\nonumber\\
\flamoi&& + \left[{\cal P}_{\iT
      \jT}(n^\qT)(\gr{n}.\gr{v}-1)+\frac{6}{5}{\cal B}_{\iT}^{\,\,\lT}\epsilon_{\jT\pT\lT}v^{\pT}+\frac{{\cal I}_\iT
  v_\jT}{2}-{\cal I}_\emptyset v_\iT v_\jT\right.\nonumber\\
\flamoi &&\left.+\left(6{\cal E}_{\iT \kT}-{\cal I}_{\iT \kT}\right)\frac{n^\kT v_\jT}{5}+\left(\frac{3}{5}{\cal E}_{\iT \jT}-\frac{{\cal I}_{\iT \jT}}{10}
      \right)(1+3\gr{v}.\gr{n})\right]^{^{\mathrm{TT}}}\,.\nonumber
\eea
This expression agrees with the results obtained for the unpolarized case~\cite{Dodelson1993,Hu1994,Bartolo2006,Maartens1999} (see details in Ref.~\cite{Pitrou2008}). The perturbative expansion
of the form~(\ref{DecL}) can then be easily read. The expression of the
collision tensor together with the Liouville operator obtained in
Eqs.~(\ref{Liouville2}) and (\ref{Liouville11}) are our key results. It is then
possible to shift to Fourier space and expand the Liouville operator and the
collision tensor either in PSTF multipoles, or in normal modes components~\cite{Hu1997} which are better
suited for numerical integration. Details of these extractions can be found in Ref.~\cite{Pitrou2008}.
Hence it offers, in principle, the possibility to integrate numerically the
Boltzmann hierarchy up to second order, though it is expected to be very time consuming, and approximate schemes should be
developed in order to focus on the dominant effects. \\
Computing the effects of non-linear evolution has now become extremely important since it has been
recently claimed that a non-zero non-Gaussianity in the CMB had been
detected~\cite{Yadav2007b}. Since the precision of the CMB measurements is
going to increase significantly in the forthcoming missions, disentangling the primordial
non-Gaussianity, which delivers valuable information on the primordial
universe, from the subsequent non-linear effects is a necessary task in order to
constrain better from inflation our theories of high energy physics.    \\
{\it Acknowledgements:} I thank J-P Uzan for many useful discussions on the
kinetic theory.


\ifcqg
\bibliographystyle{h-physrev}
\bibliography{compton}
\else
\bibliography{comptonprlsanstitres}
\fi


\end{document}